%% file: main.tex
\documentclass[runningheads]{llncs}

\usepackage[T1]{fontenc}
\usepackage{graphicx}
\usepackage{color}
\usepackage{subcaption}
\usepackage{adjustbox}
\usepackage{multicol}
\usepackage{multirow}
\usepackage{enumitem}
\usepackage{amsmath}

\usepackage[linesnumbered,ruled]{algorithm2e}
\usepackage{float}
\usepackage{algpseudocode}
\SetKwInput{KwInput}{Input}
\SetKwInput{KwOutput}{Output}
\SetKwComment{Comment}{// }{ }

\usepackage{tikz}
\usetikzlibrary{patterns}
\usetikzlibrary{decorations.pathreplacing,calc}

\newcommand{\ex}{\mathrm{expr}}
\newcommand{\size}{\mathsf{size}}
\newcommand{\height}{\mathsf{h}}

\newcommand{\grammar}{\mathcal{G}}
\newcommand{\grammarSlp}{\grammar_\mathrm{SLP}}
\newcommand{\grammarChomsky}{\grammar_\mathrm{CNF}}
\newcommand{\grammarRepair}{\grammar_\mathrm{RePair}}

\newcommand{\algName}{FRAS}

\newcommand{\leftLength}{\mathit{leftLength}}
\newcommand{\rank}{\mathrm{rank}}
\newcommand{\select}{\mathrm{select}}
\newcommand{\length}{\mathit{length}}

\begin{document}

\title{Revisiting the Folklore Algorithm for Random Access to Grammar-Compressed Strings}
\titlerunning{Folklore Random Access for SLPs}

\author{Alan M. Cleary\inst{1} \and
Joseph~Winjum\inst{2} \and
Jordan~Dood\inst{3} \and
Shunsuke~Inenaga\inst{4}}

\authorrunning{A. Cleary et al.}
\institute{National Center for Genome Resources, Santa Fe, NM, USA\\
\email{acleary@ncgr.org}\and
Montana State University, Bozeman, MT, USA\\
\email{joseph.winjum@ecat1.montana.edu}\and
Hyalite Technologies LLC, Bozeman, MT, USA\\
\email{hyalitetechnologies@gmail.com}\and
Department of Informatics, Kyushu University, Fukuoka, Japan\\
\email{inenaga.shunsuke.380@m.kyushu-u.ac.jp}}

\maketitle

\begin{abstract}
Grammar-based compression is a widely-accepted model of string compression 
that allows for efficient and direct manipulations on the compressed data.
Most, if not all, such manipulations rely on the primitive \emph{random access} queries,
a task of quickly returning the character at a specified position of the original
uncompressed string without explicit decompression. While there are advanced data structures for random access to grammar-compressed strings
that guarantee theoretical query time and space bounds,
little has been done for the \emph{practical} perspective of this important problem.
In this paper, we revisit a well-known folklore random access algorithm for grammars
in the Chomsky normal form,
modify it to work directly on general grammars,
and show that this modified version is fast and memory efficient in practice.

\end{abstract}

\section{Introduction}

\emph{Random access} on grammar-compressed strings has been used as a key primitive
in a number of efficient algorithms that directly work on compressed strings, 
including pattern matching~\cite{KarpinskiRS97,MiyazakiST00,Lifshits07,YamamotoBIT11,BilleGCSVV17}, 
compressed-string indexing~\cite{GagieGKNP12},
q-gram frequencies~\cite{GotoBIT12,GotoBIT13},
detection of palindromes and repetitions~\cite{MatsubaraIISNH09,IMSIBTNS15,INIBT15}, 
convolutions~\cite{TanakaIIBT13}, finger searches~\cite{BilleCG17},
and Lempel-Ziv factorizations in compressed space~\cite{GagieGJN24}.

Given a grammar $\grammar$ in the Chomsky normal form for a text $T$, a folklore algorithm for this problem first computes and stores the length of the string that each non-terminal derives in $O(\size(\grammar))$-time and space, where $\size(\grammar)$ denotes the total size of the productions in $\grammar$.
Then, given a position $p$ in $T$, one can climb down the corresponding path to $T[p]$ in the derivation tree for $\grammar$ in $O(\height(\grammar))$-time, where $\height(\grammar)$ denotes the height of the derivation tree of $\grammar$.
While $\height(\grammar)$ can be as small as $\Theta(\log n)$ for some highly repetitive strings of length $n$ with balanced grammars $\grammar$, $\height(\grammar)$ can be as large as $\Theta(n)$ in the worst case.
Bille et al.~\cite{BilleLRSSW15} showed how to preprocess $\grammar$ in $O(\size(\grammar))$-time and space so that later 
random access queries can be answered in $O(\log n)$-time, irrespective of $\height(\grammar)$.
Garnardi et al.~\cite{GanardiJL21} showed how to convert a given grammar $\grammar$ into another grammar $\grammar'$,
with $\size(\grammar') = O(\size(\grammar))$ and $\height(\grammar') = O(\log n)$,
that derives the same string as the original grammar $\grammar$,
thus achieving $O(\log n)$-time random access using $O(\size(\grammar))$-space.
Garnardi et al.~\cite{GanardiJL21} also presented an $O(\log n / \log \log n)$-time
random access data structure with $O(\size(\grammar)\log^\epsilon n)$-space
for any $\epsilon > 0$, by generalizing the result of Belazzougui et al.~\cite{BelazzouguiCPT15}.
This matches the cell-probe lower bound shown by Verbin and Yu~\cite{VerbinY13}, for strings that are only polynomially compressible in $n$.

While random access on grammars has been extensively studied in the theoretical perspective, as shown above, the only practical results that we are aware of are the works by Maruyama et al.~\cite{MaruyamaSPIRE13} and Gagie et al.~\cite{GagieIMNSBT20}.
However, these approaches require specific grammar encodings and only work on RePair style grammars, making them incompatible with recent grammar-based compression algorithms~\cite{NunesDCC18,FuruyaTNIBK20,ClearyDCC23}.
In fact, we are not aware of a general random access algorithm that will work on any grammar.

In this work, we revisit the folklore algorithm for random access to grammar compressed strings and show that it can be improved to use significantly less space and generalized to operate directly on any grammar.
Our experiments show that this modified folklore algorithm achieves state-of-the-art performance in both its space requirements and run-time.

\section{Preliminaries}

In this section, we define syntax and review information related this paper.
Indexes start at 1.

\subsection{Strings}
Let $\Sigma$ be an alphabet of size $\sigma$.
An element in $\Sigma^*$ is called a \emph{string}.
The length of a string $T$ is denoted by $|T|$ and $n=|T|$.
The empty string $\varepsilon$ is the string of length 0, i.e. $|\varepsilon| = 0$.
The $p$th character in a string $T$ is denoted by $T[p]$ for $1 \leq p \leq n$, and the substring of $T$ that begins at position $p$ and ends at position $q$ is denoted by $T[p..q]$ for $1 \leq p \leq q \leq n$.
For convenience, let $T[p..q] = \varepsilon$ for $p > q$.

\subsection{Grammar-based compression}
A \emph{context-free grammar} is a set of recursive rules that describe how to form strings from a language's alphabet.
A context-free grammar is called an \emph{admissible grammar} if the language it generates consists only of a single string.
\emph{Grammar-based compression} is a compression technique that computes an admissible grammar for a given string such that the computed grammar can be stored in less space than the original string.
In what follows, we will call admissible grammars simply \emph{grammars}.

Let $\grammar=\langle X,\Sigma,R,S \rangle$ be a grammar that generates $T$, where $X$ is a set of non-terminal characters, $\Sigma$ is the alphabet of $T$ (i.e. terminal characters) and is disjoint from $X$, $R$ is a finite relation in $X \times (X \cup \Sigma)^*$, and $S$ is the symbol in $X$ that should be used as the \emph{start rule} when using $\grammar$ to generate $T$.
$R$ defines the \emph{rules} of $\grammar$ as a set of $m$ productions $\{X_i \rightarrow \ex_i \mid 1 \leq i \leq m\}$ such that each $X_i$ is a non-terminal in $X$ and $\ex_i$ is a non-empty sequence from $(\Sigma \cup \{X_1, \ldots, X_{i-1}\})^+$.
The \emph{size} of grammar $\grammar$ is the total length of the right-hand sides of the productions and is denoted $\size(\grammar) = \sum_{i = 1}^{m}|\ex_i|$.
We say that a non-terminal $X_i$ \emph{represents} a string $w$ if $w$ is the (unique) string that $X_i$ derives.
We only consider grammars with no useless rules and symbols, unless stated otherwise.
$\height(\grammar)$ denotes the height of the derivation tree of grammar $\grammar$.

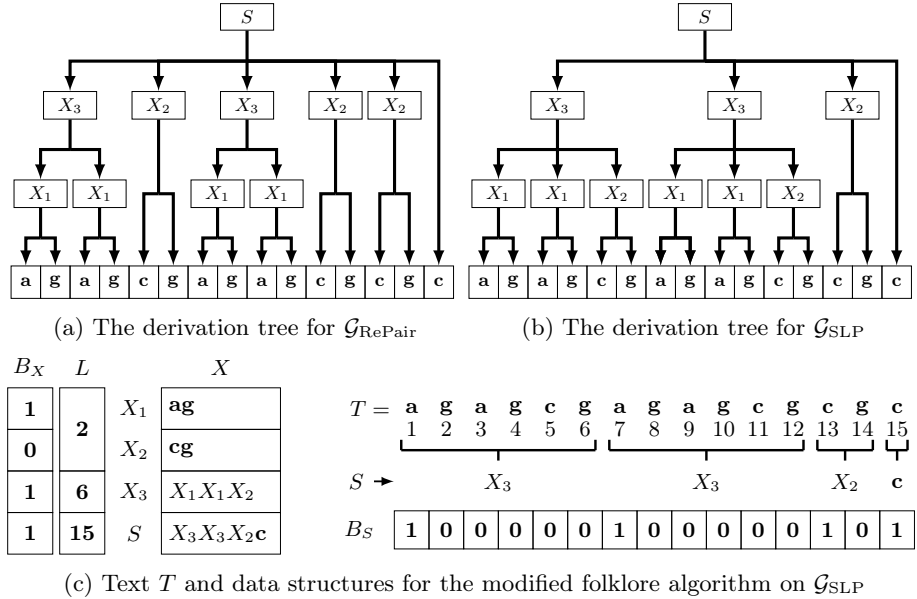
\begin{figure}[h!]
\begin{center}
    \begin{subfigure}[t]{0.49\textwidth}
        \centering
        \begin{adjustbox}{max width=1\textwidth}
            \input{Figures/Panel_A.tex}
        \end{adjustbox}
        \caption{The derivation tree for $\grammarRepair$}
    \end{subfigure}
    \begin{subfigure}[t]{0.49\textwidth}
        \centering
        \begin{adjustbox}{max width=1.0\textwidth}
            \input{Figures/Panel_B.tex}
        \end{adjustbox}
        \caption{The derivation tree for $\grammarSlp$}
    \end{subfigure} \\
    \begin{subfigure}[t]{\textwidth}
        \centering
        \begin{adjustbox}{max width=1\textwidth}
            \input{Figures/Panel_C.tex}
        \end{adjustbox}
        \caption{Text $T$ and data structures for the modified folklore algorithm on $\grammarSlp$}
    \end{subfigure}
\end{center}
\caption{Example grammars and data structures for the modified folklore algorithm on string $T = \text{agagcgagagcgcgc}$. (a) and (b) depict the derivation trees for a RePair grammar $\grammarRepair$ and an SLP grammar $\grammarSlp$, respectively. (c) depicts the data structures used by our modified version of the folklore random access algorithm. Note that the subscript $i$ for each non-terminal $X_i$ conveys the array index the non-terminal in the array of arrays representation of the grammars. For simplicity, terminal characters are used directly without a proxy non-terminal character.}\label{fig:example}
\end {figure}

In this work we will discuss three types of grammars: \emph{straight-line programs (SLPs)}, \emph{Chomsky normal form (CNF) grammars}, and \emph{RePair grammars}.
An SLP $\grammarSlp$ is simply an admissible grammar.
A CNF grammar $\grammarChomsky$ is an SLP in the Chomsky normal form, i.e. every rule (including start rule $S$) is of the form $X_i \rightarrow c$~($c \in \Sigma$) or $X_i \rightarrow X_\ell X_r$~($\ell, r < i$)\footnote{While the term SLP is often used for grammar-compression in the Chomsky normal form, in this paper, for clarity, we use SLP to denote general admissible grammars and CNF to denote grammars in the Chomsky normal form.}.
A RePair grammar $\grammarRepair$, proposed in~\cite{LarssonIEEE00}, is a CNF grammar in which the start rule can have arbitrarily many symbols in its righthand side, i.e. $S \rightarrow \ex$ with $\ex \in ((X \setminus S) \cup \Sigma)^+$.
Any SLP can be converted to an equivalent CNF grammar~\cite{LohreyGCC12} and any CNF grammar can be converted to an equivalent balanced CNF grammar with height $O(\log n)$~\cite{GanardiJL21}, where $n$ is the length of the uncompressed string.
See Figure~\ref{fig:example}~(a) and~(b) for example SLP and RePair grammars.

Let $\grammar$ be a grammar that represents a string $T$ of length $n$.
A \emph{random access query} on grammar $\grammar$ is, given a query position $p$~($1 \leq p \leq n$),
return the $p$th character $T[p]$ in the uncompressed string $T$.
The \emph{random access problem on grammar-compression} is to preprocess a given grammar $\grammar$ 
and build a space-efficient (i.e. compressed) data structure on $\grammar$ that
can quickly return the desired character $T[p]$ for query positions $p$.
In practice, random access queries can be for entire substrings $T[p..q]$, where $1 \leq p \leq q \leq n$.
Our experiments include results for substring queries but our algorithm descriptions only consider the single character version of the problem.
This is because in our approach only the first character of the substring needs to be located in $\grammar$'s derivation tree; the rest of the substring can be generated by simply traversing the derivation tree from that location.

\section{Algorithms}
\label{sec:algorithms}

In this section, we present novel algorithms for random access to grammar-compressed strings.
As before, indexes start at 1.

\begin{algorithm}[h]
\DontPrintSemicolon

  \KwData{Grammar $\grammarChomsky$, i.e. an array of integer arrays}
  \KwData{Start rule $S$, i.e. an index in $\grammarChomsky$}
  \KwData{Length lookup table $A$}
  \KwInput{Position $p$ in $T$}
  \KwOutput{Character $T[p]$}

  $X_i \gets S$\;
  \While{$X_i$ is a non-terminal ($X_i \rightarrow X_\ell X_r$)}{ 
        $\leftLength \gets A[X_i]$\;
        \eIf{$p \leq \leftLength$}{
            $X_i \gets \grammarChomsky[X_i]_\ell$ \Comment{$X_\ell$ denotes the first character in $X_i$'s array}
        }{
            $p \gets p - \leftLength$\;
            $X_i \gets \grammarChomsky[X_i]_r$ \Comment{$X_r$ denotes the second character in $X_i$'s array}
        } 
    } 
    \Return{$X_i$}

\caption{Folklore Random Access}\label{alg:folklore}
\end{algorithm}

\begin{algorithm}[h]
\DontPrintSemicolon

  \KwData{Ordered grammar $\grammarSlp$, i.e. an array of integer arrays}
  \KwData{Start rule $S$, i.e. an index in $\grammarSlp$}
  \KwData{Bitvector $B_X$}
  \KwData{Bitvector $B_S$}
  \KwData{Unique length array $L$}
  \KwInput{Position $p$ in $T$}
  \KwOutput{Character $T[p]$}

  $r \gets \rank(B_S, p)$\;
  $X_i \gets \grammarSlp[S][r]$\;
  $p \gets p - \select(B_S, r)$\;

  \While{$X_i$ is a non-terminal ($X_i \rightarrow \ex_i$ with $\ex_i \in (\Sigma \cup \{X_1, \ldots, X_{i-1}\})^+$)}{
      \For{$X_j$ in $\grammarSlp[X_i]$}{
          $r \gets \rank(B_X, X_j)$\;
          $\length \gets L[r]$\;
          \eIf{$p \leq \length$}{
              $X_i \gets X_j$\;
              break\;
          }{
              $p \gets p - \length$\;
          }
      }
  }
  
  \Return{$X_i$}\;

\caption{Modified Folklore Random Access}\label{alg:modified-folklore}
\end{algorithm}

\subsection{The Folklore Algorithm}

The folklore algorithm for random access to grammar-compressed strings is for CNF grammars.
Given a grammar $\grammarChomsky$, the algorithm works by first computing an array $A$ that stores the length of the string that each non-terminal represents.
Then, given a position $p$ in $T$, the corresponding path to $T[p]$ in the derivation tree is followed by looking up the string length of each non-terminal's left and right characters in $A$ to determine which character contains the position, requiring $O(\log n)$-time when the grammar is balanced and $O(\height(\grammarChomsky))$-time otherwise.
See Algorithm~\ref{alg:folklore} for details.

While this algorithm is fast both asymptotically and in practice, it requires much additional space.
For instance, if the algorithm is given a non-CNF grammar $\grammar$, then an equivalent CNF grammar $\grammarChomsky$ must be computed, which may require introducing useless rules that increase the size of the grammar.
Additionally, using the na\"ive array of arrays grammar encoding~\cite{TabeiCPM13} where each non-terminal character is an array index, the folklore algorithm requires:
\begin{enumerate}[label=(\alph*)]
    \item 
    $2m\lg(m + \sigma)$ bits to represent the grammar and\label{folklore:grammar}
    \item 
    $m\lg(n)$ bits to represent the rule string lengths,\label{folklore:lengths}
\end{enumerate}
where $m$ denotes the number of non-terminals in $\grammarChomsky$,
in which there are $\sigma$ rules of form $X_i \rightarrow c$~($c \in \Sigma$).
Note that the bits to represent the rule string lengths require half as much space as the grammar itself!

\subsection{Modified Folklore Algorithm}

In our modified folklore algorithm to follow,
the non-terminals are required to be sorted (and subsequently numbered) in increasing order of their expansion lengths.
Given a grammar $\grammar$ with $m$ non-terminals,
we can simply sort them in $O(m \log m)$-time and renumber them in $O(\size(\grammar))$-time
with $O(m \log(n))$ bits of working space by any suitable comparison-based sorting algorithm.
The $O(m \log(n))$ bits of information is discarded after this preprocessing.
If the non-terminals in $\grammar$ are already sorted, this step can be skipped.

To improve the folklore algorithm, we first observe that it can be easily extended to work on any SLP.
Specifically, given an SLP $\grammarSlp$, the length of the string each non-terminal represents is computed and stored, as before.
Then the position in $T$ of each non-terminal character in start rule $S$ is computed and stored as well.
Now, given a position $p$ in $T$, the algorithm will first look up the character in the start rule that contains $p$ and then descend the derivation tree from this character.
However, since each rule in an SLP can have arbitrarily many symbols in its righthand side, instead of checking the left and right characters to determine which character contains the query position, a rule's characters are iterated until the character containing the position is found.

Using the na\"ive array of arrays grammar encoding, this extended version of the folklore algorithm requires:
\begin{enumerate}[label=(\alph*)]
    \setcounter{enumi}{2}
    \item $\size(\grammarSlp)\lg(m + \sigma)$ bits to represent the grammar and\label{folklore-slp:grammar}
    \item $(|S| + m)\lg(n)$ bits to represent the rule string lengths and start character positions.\label{folklore-slp:lengths}
\end{enumerate}
Note that when $\grammarSlp$ is a CNF grammar, \ref{folklore-slp:grammar} is equivalent to~\ref{folklore:grammar}.

We observe that the following changes can be made to significantly reduce the space requirements of \ref{folklore-slp:lengths}:
\begin{enumerate}
    \item Using the na\"ive grammar representation of \ref{folklore-slp:grammar} where each non-terminal character is an array index, order (and subsequently renumber) the rules by string length -- shortest to longest -- with the start rule $S$ last.
    \item Create a sorted array $L$ of the unique rule string lengths.
    \item Create a length $m$ bitvector $B_X$ with a bit $i$ set for the first rule $X_i$ of each string length in the grammar array.
    \item Create a length $n$ bitvector $B_S$ with a bit $p$ set at the text position of each character in start rule $S$.
\end{enumerate}
The modified folklore algorithm can then be used by first looking up the start rule character via a paired \emph{rank-select} query on $B_S$.
The algorithm can then descend the derivation tree by using \emph{rank} queries on $B_X$ to look up each rule's string length in $L$.
See Algorithm~\ref{alg:modified-folklore} for details.

Using a standard bitvector the \emph{rank} and \emph{select} operations can be done in $O(1)$-time~\cite{VignaAlgorithms08,Clark1Dissertation97}.
However, a standard bitvector requires $64 \lceil |B| / 64 + 1 \rceil$ bits of space, where $|B|$ is the length of the bitvector.
This is impractical since $|B_S|=n$.

To minimize the space required by bitvectors $B_S$ and $B_X$, we propose using the sparse bitvector~\cite{OkanoharaALENEX07}.
This bitvector answers \emph{rank} queries in $O(\log{\frac{|B|}{b}})$-time and select queries in $O(1)$-time while requiring no more than $b (2 + \log{\frac{|B|}{b}})$ bits of space, where $b$ is the number of set bits.
Using the sparse bitvector, the modified folklore algorithm requires
no more than $|S| (2 + \log{\frac{n}{|S|}}) + |L| (2 + \log{\frac{m}{|L|}}) + |L| \lg |L| $ bits to represent the rule string lengths and start character positions.

We observe that the modified folklore algorithm is effectively equivalent to the original folklore algorithm when given a CNF grammar.
Since any SLP can be converted to an equivalent balanced CNF grammar~\cite{LohreyGCC12,GanardiJL21}, this implies that when using standard bitvectors the modified folklore algorithm runs in $O(\log n)$-time, and when using sparse bitvectors it runs in $O(\log{\frac{n}{|S|}} + \log n \log{\frac{m}{|L|}})$-time.
Without converting to the Chomsky normal form or balancing the grammar, the worst-case run-time is $O(n)$.
However, algorithms that produce RePair grammars tend to create balanced grammars, and so the expected run-time on RePair grammars is $O(\log n)$.
Although the nature of SLPs in general is less predictable, the grammars of the MR-RePair algorithm we use in our experiments have been shown to be isomorphic to grammars produced by the actual RePair algorithm~\cite{FuruyaTNIBK20}, so we expect these grammars to have an $O(\log n)$ run-time as well, despite having to iterate rules' characters when descending the derivation tree.

\section{Results}

In this section, we describe our implementation and experimental results.

\subsection{Implementation}

Our implementation of the modified folklore algorithm is called \emph{FRAS} - Folklore Random Access for SLPs.
We implemented FRAS in C++.
The bitvectors and their respective \emph{rank} and \emph{select} data structures were implemented using the Succinct Data Structure Library (SDSL)~\cite{GogAlgorithms14}.
We used the na\"ive array of arrays encoding to represent grammars~\cite{TabeiCPM13} and implemented the sparse bitvector variation of the folklore algorithm described in Section~\ref{sec:algorithms}.
The source code is available at \url{https://github.com/alancleary/FRAS}.

\subsection{Experiments}
\label{sec:experiments}

We performed experiments on two corpora of data: the \emph{Pizza\&Chili} corpus\footnote{\url{https://pizzachili.dcc.uchile.cl/}} and a collection of pangenomes.
For each corpus, we generated grammars and benchmarked our modified folklore algorithm (\algName) against the original folklore algorithm (Folklore), the algorithm of~\cite{MaruyamaSPIRE13} (FOLCA), and the algorithm of~\cite{GagieIMNSBT20} (ShapedSLP\footnote{Note that in \cite{GagieIMNSBT20} ``SLP'' refers to RePair grammars, thus ShapedSLP is only compatible with RePair grammars.}), measuring encoding size and random access run-time.
For each grammar generated, we accessed substrings of length 1, 10, 100, and 1,000 at pseudo-random positions\footnote{Pseudo-random numbers were generated on a uniform distribution using the xoroshiro128+ generator~\cite{BlackmanACM2021}. The generator was seeded so that the same numbers were used by every algorithm.}
We performed this procedure 10,000 times for each substring length and computed the average run-time.
See the following subsections for details.

Experiments were run on a server with two AMD EPYC 7543 32-Core 2.8GHz (3.7GHz max boost) processors and 2TB of 8-channel DDR4 3200MHz memory running CentOS Stream release 9.
Note that this server is excessively overpowered for these experiments and was used for the purpose of stability and accuracy of measurement.
Similar results can be achieved on a consumer laptop.

\subsubsection{Pizza\&Chili}

For the Pizza\&Chili corpus, we generated grammars for all data sets from the \emph{real} and \emph{artificial} collections.
For each data set, we generated grammars using Gonzalo Navarro's implementation\footnote{\url{https://users.dcc.uchile.cl/~gnavarro/software/repair.tgz}} of RePair~\cite{LarssonIEEE00} and Isamu Furuya's implementation\footnote{\url{https://github.com/izflare/MR-RePair}} of MR-RePair\cite{FuruyaTNIBK20}.
MR-RePair generates SLPs that are isomorphic to RePair grammars.
We included these grammars to test our hypothesis that the run-times of our \algName \ algorithm should be approximately equivalent on an SLP as on an equivalent RePair grammar.
The Folklore, FOLCA, and ShapedSLP algorithms were not benchmarked on MR-RePair SLPs as they only work on RePair grammars.
See Table~\ref{table:dataPizza} in Appendix~\ref{appendix} for information about the Pizza\&Chili data sets and their respective grammars.
The space used by each algorithm is listed in Table~\ref{table:pizzachili-sizes} and the run-times of each algorithm are listed in Table~\ref{table:pizzachili-runtimes}.

We found that FRAS consistently used less space than Folklore, especially on the MR-RePair grammars.
However, FRAS was also consistently slower than Folklore.
This is expected as both FOLCA and ShapedSLP also use less space than Folklore but are slower.
FRAS, however, consistently used more space than FOLCA and ShapedSLP but was also much faster.
Moreover, FRAS was faster than FOLCA and ShapedSLP on every data set and query size, and it was the only algorithm of the three to achieve sub-microsecond run-times.

Interestingly, FRAS on MR-RePair grammars did not use much more space than FOLCA and ShapedSLP but it was the fastest of all the algorithms, excluding Folklore.
This suggests that although the size of the rules in MR-RePair grammars is unbounded FRAS is acheiving the expected run-time and is faster than the FRAS on RePair grammars simply because MR-RePair grammars are typically smaller.

\subsubsection{Pangenomes}

A \emph{pangenome} is a collection of genomes from the same species.
The ability to efficiently store and access these data at scale is an important problem in bioinformatics~\cite{PangenomeBriefings16}.
Grammar-based compression is particularly well suited to compressing these data as the size of the grammars scales relative to information content, rather than input size~\cite{NavarroACM21}.
To demonstrate the practicality of our algorithm, we generated grammars for three pangenomes:
the 12 yeast assemblies from the Yeast Population Reference Panel (YPRP)~\cite{YueNature2017};
the 25 Maize assembles from the nested association mapping (NAM) population~\cite{HuffordScience21};
and 1000 copies of Human chromosome 19 (c1000) used by Gagie et al. in~\cite{GagieIMNSBT20}.
Grammars were generated using BigRePair as it is currently the only grammar-based compression algorithm that can generate grammars for the NAM and c1000 data sets~\cite{GagieSPIRE19}.
See Table~\ref{table:dataPan} in Appendix~\ref{appendix} for information about these data sets and their respective grammars.
The space used by each algorithm is listed in Table~\ref{table:pangenome-sizes} and the run-times of each algorithm are listed in Table~\ref{table:pangenome-runtimes}.

As with the Pizza\&Chili corpus, we found that FRAS consistently used less space than Folklore and was consistently slower.
And, again, FRAS consistently used more space than FOLCA and ShapedSLP but was also much faster, beating FOLCA and ShapedSLP on every data set and query size.

\begin{table}[b]
    \centering
    \caption{The space used by the random access algorithms benchmarked on the \emph{Pizze\&Chili} corpus. \textbf{Data Set} is the names of the data sets and what collection they belong to. The \textbf{MR-\algName}, \textbf{\algName}, \textbf{Folklore}, \textbf{FOLCA}, and \textbf{ShapedSLP} columns are the space used by each algorithm, where \textbf{MR-\algName} is \textbf{\algName} run on MR-RePair grammars; all other results are on RePair grammars. All space is in megabytes.}
    \label{table:pizzachili-sizes}
    \par\vspace{\baselineskip}
    \begin{tabular}{|c|l|c|c|c|c|c|}
        \hline
        \multicolumn{2}{|c|}{\textbf{Data Set}} & \textbf{MR-\algName} & \textbf{\algName} & \textbf{Folklore} & \textbf{FOLCA} & \textbf{ShapedSLP} \\
        \hline
        \multirow{9}*{\rotatebox[origin=c]{90}{\emph{real}}}
         & \emph{Escherichia\_Coli} & 17.2   & 22.9   & 41.1   & 13.8   & 14.0 \\
         & \emph{cere}              & 16.3   & 23.7   & 37.4   & 13.1   & 14.0 \\
         & \emph{coreutils}         & 9.8    & 16.6   & 22.7   & 7.9    & 9.0  \\
         & \emph{einstein.de.txt}   & 0.4    & 0.5    & 0.7    & 0.3    & 0.3  \\
         & \emph{einstein.en.txt}   & 0.9    & 1.2    & 1.9    & 0.7    & 0.6  \\
         & \emph{influenza}         & 8.0    & 8.8    & 17.6   & 6.1    & 5.5  \\
         & \emph{kernel}            & 5.6    & 9.2    & 13.0   & 4.5    & 5.0  \\
         & \emph{para}              & 21.3   & 29.5   & 49.3   & 17.4   & 18.2 \\
         & \emph{world\_leaders}    & 1.7    & 2.1    & 3.4    & 1.1    & 1.1  \\
        \hline
        \multirow{3}*{\rotatebox[origin=c]{90}{\emph{artificial}}}
         & \emph{fib41}             & $<0.1$ & $<0.1$ & $<0.1$ & $<0.1$ & $<0.1$ \\
         & \emph{rs.13}             & $<0.1$ & $<0.1$ & $<0.1$ & $<0.1$ & $<0.1$ \\
         & \emph{tm29}              & $<0.1$ & $<0.1$ & $<0.1$ & $<0.1$ & $<0.1$ \\
        \hline
    \end{tabular}
\end{table}

\newcommand{\algHeader}[1]{\multicolumn{4}{c|}{\textbf{#1}}}
\newcommand{\queryHeaders}{\textbf{1} & \textbf{10} & \textbf{100} & \textbf{1,000}}
\begin{table}[b]
    \caption{Random access run-times for grammars built on the \emph{Pizze\&Chili} corpus. \textbf{Data Set} is the names of the data sets and what collection they belong to. The \textbf{MR-\algName}, \textbf{\algName}, \textbf{Folklore}, \textbf{FOLCA}, and \textbf{ShapedSLP} columns are the algorithms benchmarked and their query run-times, where \textbf{MR-\algName} is \textbf{\algName} run on MR-RePair grammars; all other results are on RePair grammars. For the run-times, \textbf{1}, \textbf{10}, \textbf{100}, and \textbf{1,000} are the lengths of the substrings queried. All run-time are in microseconds.}
    \label{table:pizzachili-runtimes}
    \par\vspace{\baselineskip}
    \hspace{-3.3cm}
    \begin{tabular}{|c|l|c|c|c|c|c|c|c|c|c|c|c|c|c|c|c|c|c|c|c|c|}
        \hline
        \multicolumn{2}{|c|}{\multirow{2}*{\textbf{Data Set}}} & \algHeader{MR-\algName} & \algHeader{\algName} & \algHeader{Folklore} & \algHeader{FOLCA} & \algHeader{ShapedSLP} \\ 
        \cline{3-22}
        \multicolumn{2}{|c|}{} & \queryHeaders & \queryHeaders & \queryHeaders & \queryHeaders & \queryHeaders \\
        \hline
        \multirow{9}*{\rotatebox[origin=c]{90}{\emph{real}}}
         & \emph{Escherichia\_Coli} & 1.8  & 1.9  & 3.4  & 17.7 & 5.8  & 6.0  & 8.4  & 29.2 & 1.1 & 1.2 & 2.2 & 12.0 & 25.0  & 26.3  & 42.0  & 191.2 & 21.0  & 22.7  & 42.5  & 238.7 \\
         & \emph{cere}              & 1.3  & 1.5  & 3.0  & 16.2 & 4.4  & 4.7  & 7.2  & 28.5 & 1.0 & 1.1 & 2.1 & 10.9 & 17.7  & 19.4  & 34.7  & 182.3 & 14.2  & 16.3  & 36.5  & 235.8 \\
         & \emph{coreutils}         & 17.5 & 17.8 & 19.5 & 34.5 & 42.2 & 42.7 & 46.3 & 69.1 & 6.4 & 7.2 & 9.5 & 20.9 & 156.0 & 158.3 & 180.7 & 355.4 & 242.9 & 247.2 & 276.7 & 491.9 \\
         & \emph{einstein.de.txt}   & 0.8  & 1.0  & 2.4  & 15.2 & 1.4  & 1.6  & 3.4  & 19.0 & 0.5 & 0.7 & 1.6 & 10.0 & 10.6  & 12.3  & 28.1  & 176.7 & 6.6   & 8.6   & 27.8  & 217.6 \\
         & \emph{einstein.en.txt}   & 1.0  & 1.2  & 2.8  & 16.5 & 1.3  & 1.5  & 3.4  & 19.2 & 0.7 & 0.8 & 1.8 & 10.2 & 11.8  & 13.6  & 30.2  & 184.2 & 6.2   & 8.3   & 27.7  & 217.9 \\
         & \emph{influenza}         & 0.3  & 0.5  & 2.2  & 17.1 & 0.5  & 0.6  & 2.3  & 17.0 & 0.6 & 0.7 & 1.6 & 10.4 & 10.6  & 12.3  & 27.3  & 172.1 & 3.0   & 5.0   & 24.1  & 213.6 \\
         & \emph{kernel}            & 5.1  & 5.3  & 7.0  & 21.6 & 15.2 & 15.9 & 18.1 & 36.0 & 1.9 & 2.2 & 3.4 & 13.3 & 51.7  & 54.6  & 72.3  & 241.3 & 60.5  & 64.3  & 85.7  & 289.7 \\
         & \emph{para}              & 0.9  & 1.1  & 2.6  & 17.4 & 1.8  & 2.2  & 4.8  & 27.4 & 0.8 & 1.0 & 2.1 & 11.9 & 13.2  & 14.9  & 30.7  & 182.3 & 5.8   & 7.7   & 27.0  & 218.8 \\
         & \emph{world\_leaders}    & 0.5  & 0.7  & 1.9  & 13.0 & 1.1  & 1.3  & 2.7  & 14.5 & 0.6 & 0.7 & 1.3 & 7.6  & 11.8  & 13.3  & 28.6  & 166.7 & 5.7   & 7.7   & 26.7  & 214.0 \\
        \hline
        \multirow{3}*{\rotatebox[origin=c]{90}{\emph{artificial}}}
         & \emph{fib41}             & 0.9  & 1.0  & 1.2  & 7.4  & 1.0  & 1.0  & 1.2  & 7.4  & 0.2 & 0.3 & 0.6 & 3.6  & 2.6   & 3.3   & 10.4  & 81.3  & 4.2   & 5.7   & 20.8  & 172.1 \\
         & \emph{rs.13}             & 0.9  & 1.0  & 1.3  & 7.9  & 0.9  & 1.0  & 1.2  & 7.4  & 0.3 & 0.3 & 0.6 & 3.8  & 3.1   & 3.8   & 11.2  & 85.8  & 4.7   & 6.2   & 22.0  & 179.1 \\
         & \emph{tm29}              & 0.9  & 0.9  & 1.2  & 7.0  & 0.8  & 0.9  & 1.3  & 7.6  & 0.2 & 0.3 & 0.6 & 3.7  & 2.8   & 3.6   & 10.6  & 80.9  & 4.7   & 6.3   & 22.5  & 183.1 \\
        \hline
    \end{tabular}
\end{table}

\begin{table}[b]
    \centering
    \caption{The space used by the random access algorithms benchmarked on the \emph{Pangenome} corpus. \textbf{Data Set} is the names of the data sets and the \textbf{\algName}, \textbf{Folklore}, \textbf{FOLCA}, and \textbf{ShapedSLP} columns are the space used by each algorithm. All results are on BigRePair grammars. All space is in megabytes.}
    \label{table:pangenome-sizes}
    \par\vspace{\baselineskip}
    \begin{tabular}{|l|c|c|c|c|}
        \hline
        \textbf{Data Set} & \textbf{\algName} & \textbf{Folklore} & \textbf{FOLCA} & \textbf{ShapedSLP} \\
        \hline
        \emph{YPRP}  & 49.9   & 75.4   & 25.9   & 31.4   \\
        \emph{Maize} & 3779.6 & 6400.4 & 2529.8 & 3020.2 \\
        \emph{c1000} & 122.3  & 217.4  & 86.5   & 80.6   \\
        \hline
    \end{tabular}
\end{table}

\begin{table}[b]
    \centering
    \caption{Random access run-times for grammars built on the \emph{Pangenome} corpus. \textbf{Data Set} is the names of the data sets and the \textbf{\algName}, \textbf{Folklore}, \textbf{FOLCA}, and \textbf{ShapedSLP} columns are the algorithms benchmarked and their query run-times. For the run-times, \textbf{1}, \textbf{10}, \textbf{100}, and \textbf{1,000} are the lengths of the substrings queried. All results are on BigRePair grammars. All run-time are in microseconds.}
    \label{table:pangenome-runtimes}
    \par\vspace{\baselineskip}
    \begin{tabular}{|l| c|c|c|c| c|c|c|c| c|c|c|c| c|c|c|c|}
        \hline
        \multirow{2}*{\textbf{Data Set}} & \algHeader{\algName} & \algHeader{Folklore} & \algHeader{FOLCA} & \algHeader{ShapedSLP} \\ 
        \cline{2-17}
         & \queryHeaders & \queryHeaders & \queryHeaders & \queryHeaders \\
        \hline
        \emph{YPRP}  & 1.8 & 2.2 & 5.7 & 37.4 & 0.9 & 1.1 & 2.5 & 14.7 & 11.5 & 13.3 & 29.6 & 187.1 & 2.9 & 4.9  & 24.9 & 220.1 \\
        \emph{Maize} & 3.9 & 4.4 & 8.5 & 44.5 & 3.1 & 3.6 & 5.7 & 24.0 & 22.3 & 24.3 & 44.3 & 233.6 & 9.0 & 11.2 & 35.1 & 269.5 \\
        \emph{c1000} & 3.2 & 3.6 & 7.4 & 41.0 & 2.5 & 2.7 & 4.8 & 21.2 & 16.4 & 18.3 & 34.8 & 192.2 & 6.9 & 9.2  & 31.1 & 242.2 \\
        \hline
    \end{tabular}
\end{table}

\section{Conclusion}

In this work, we showed how the folklore algorithm for random access to grammar-compressed strings can be modified to work on any grammar and achieve good space and run-time performance in practice.
We believe this is the first random access algorithm for grammar-compressed strings that works directly on any grammar, thus further enhancing the usefulness of the algorithm.
In future work we would like to further improve the modified folklore algorithm by representing the grammar in a manner that uses less space with minimal effect on run-time performance.
 
\begin{credits}
\subsubsection{\ackname} 
The work of Alan M. Cleary, Joseph Winjum, and Jordan Dood was support by NSF award number 2105391.
The work of Shunsuke Inenaga was supported by JSPS KAKENHI grant numbers JP	20H05964, JP23K24808, JP23K18466.
We would like to thank the authors of~\cite{GagieIMNSBT20} for sharing the c1000 data set with us.
You saved us much time and computation.

\end{credits}

\clearpage

\bibliographystyle{splncs04}
\bibliography{main}

\clearpage
\appendix
\section{Appendix}\label{appendix}

\newcommand{\grammarHeaders}{\textbf{Rules} & \textbf{Depth} & \textbf{Start} & \textbf{Size}}
\renewcommand{\arraystretch}{1.1}
\begin{table}
    \caption{Data sets used from the \emph{Pizze\&Chili} corpus. \textbf{Data Set} is the names of the data sets and what collection they belong to, \textbf{Size} is the number of characters in each data set, and \textbf{MR-RePair} and \textbf{RePair} are information about the grammars generated for these data sets. For the grammars, \textbf{Rules} is the number of rules in the grammars, \textbf{Depth} is the maximum depth of the grammars, \textbf{Start} is the size of the start rules, and \textbf{Size} is the total lengths of the right-hand sides of the rules in each grammar, excluding the start rule.} 
    \label{table:dataPizza}
    \par\vspace{\baselineskip}
    \hspace{-1.5cm}
    \begin{tabular}{|c|l|c|c|c|c|c|c|c|c|c|}
        \hline
        \multicolumn{2}{|c|}{\multirow{2}*{\textbf{Data Set}}} & \multirow{2}*{\textbf{Size}} & \multicolumn{4}{c|}{\textbf{MR-RePair}} & \multicolumn{4}{c|}{\textbf{RePair}} \\ 
        \cline{4-11}
        \multicolumn{2}{|c|}{} & & \grammarHeaders & \grammarHeaders \\
        \hline
        \multirow{9}*{\rotatebox[origin=c]{90}{\emph{real}}}
         & \emph{Escherichia\_Coli} & 112,689,515 & 712,228   & 23 & 712,484   & 1,595,881 & 2,012,087 & 3,279  & 1,601,482 & 4,024,174 \\
         & \emph{cere}              & 461,286,644 & 836,956   & 29 & 648,659   & 3,392,707 & 2,561,292 & 1,359  & 655,298   & 5,122,584 \\
         & \emph{coreutils}         & 205,281,778 & 437,054   & 30 & 153,775   & 2,270,187 & 1,821,734 & 43,728 & 153,346   & 3,643,468 \\
         & \emph{einstein.de.txt}   & 92,758,441  & 21,787    & 42 & 12,683    & 71,709    & 49,949    & 269    & 12,665    & 99,898    \\
         & \emph{einstein.en.txt}   & 467,626,544 & 49,565    & 48 & 62,591    & 150,233   & 100,611   & 1,355  & 62,473    & 201,222   \\
         & \emph{influenza}         & 154,808,555 & 429,027   & 28 & 897,657   & 1,088,872 & 643,587   & 366    & 886,836   & 1,287,174 \\
         & \emph{kernel}            & 257,961,616 & 246,596   & 34 & 69,537    & 1,304,343 & 1,057,914 & 5,822  & 69,427    & 2,115,828 \\
         & \emph{para}              & 429,265,758 & 1,079,287 & 30 & 1,134,361 & 4,157,167 & 3,093,873 & 487    & 1,147,650 & 6,187,746 \\
         & \emph{world\_leaders}    & 46,968,181  & 100,293   & 30 & 98,397    & 309,222   & 206,508   & 463    & 94,327    & 413,016   \\
        \hline
        \multirow{3}*{\rotatebox[origin=c]{90}{\emph{artificial}}}
         & \emph{fib41}             & 267,914,296 & 38        & 40 & 3         & 76        & 38        & 40     & 3         & 76        \\
         & \emph{rs.13}             & 216,747,218 & 55        & 45 & 121       & 26        & 66        & 47     & 24        & 132       \\
         & \emph{tm29}              & 268,435,456 & 51        & 29 & 6         & 126       & 75        & 45     & 6         & 150       \\
        \hline
    \end{tabular}
\end{table}

\vspace{\baselineskip}

\begin{table}[h!]
    \caption{Data sets used from the \emph{Pangenome} corpus. \textbf{Data Set} is the names of the data sets, \textbf{Size} is the number of characters in each data set, and \textbf{BigRePair} is information about the grammars generated for these data sets. For the grammars, \textbf{Rules} is the number of rules in the grammars, \textbf{Depth} is the maximum depth of the grammars, \textbf{Start} is the size of the start rules, and \textbf{Size} is the total lengths of the right-hand sides of the rules in each grammar, excluding the start rule.}
    \label{table:dataPan}
    \par\vspace{\baselineskip}
    \hspace{1cm}
    \begin{tabular}{|l|c|c|c|c|c|}
        \hline
        \multirow{2}*{\textbf{Data Set}} & \multirow{2}*{\textbf{Size}} & \multicolumn{4}{c|}{\textbf{BigRePair}} \\ 
        \cline{3-6}
        & & \grammarHeaders \\
        \hline
        \emph{YPRP}  & 143,169,450    & 5,911,887   & 37 & 642,828    & 11,823,774  \\
        \emph{Maize} & 55,270,577,570 & 432,651,138 & 47 & 79,378,700 & 865,302,276 \\
        \emph{c1000} & 59,125,115,010 & 12,898,128  & 45 & 4,495,360  & 21,300,896  \\
        \hline
    \end{tabular}
\end{table}

\end{document}

%% file: Figures/Panel_A.tex

\begin{tikzpicture}

    \newcommand{\ptreelzx}{-2.5}
    \newcommand{\ptreelzy}{0}
    \newcommand{\ptreelfx}{-6.25}
    \newcommand{\ptreelfy}{-4.5}
    \newcommand{\ptreelflx}{-6}
    \newcommand{\ptreelfly}{-1.5}
    \newcommand{\ptreelslx}{-6}
    \newcommand{\ptreelsly}{-3}
    
    \node[draw, rectangle, minimum width=0.9cm, minimum height=0.45cm] (a) at (\ptreelzx, \ptreelzy) {\textbf{$S$}};
    \node[inner sep=0pt, outer sep=0pt, minimum size=0pt] (10) at (\ptreelzx, \ptreelzy - 0.75) {};

    \node[draw, rectangle, minimum width=0.5cm, minimum height=0.55cm] (b) at (\ptreelfx + 0, \ptreelfy) {\textbf{a}};
    \node[draw, rectangle, minimum width=0.5cm, minimum height=0.55cm] (c) at (\ptreelfx + 0.5, \ptreelfy) {\textbf{g}};
    \node[draw, rectangle, minimum width=0.5cm, minimum height=0.55cm] (d) at (\ptreelfx + 1, \ptreelfy) {\textbf{a}};
    \node[draw, rectangle, minimum width=0.5cm, minimum height=0.55cm] (dd) at (\ptreelfx + 1.5, \ptreelfy) {\textbf{g}};
    \node[draw, rectangle, minimum width=0.5cm, minimum height=0.55cm] (e) at (\ptreelfx + 2, \ptreelfy) {\textbf{c}};
    \node[draw, rectangle, minimum width=0.5cm, minimum height=0.55cm] (f) at (\ptreelfx + 2.5, \ptreelfy) {\textbf{g}};
    \node[draw, rectangle, minimum width=0.5cm, minimum height=0.55cm] (g) at (\ptreelfx + 3, \ptreelfy) {\textbf{a}};
    \node[draw, rectangle, minimum width=0.5cm, minimum height=0.55cm] (h) at (\ptreelfx + 3.5, \ptreelfy) {\textbf{g}};
    \node[draw, rectangle, minimum width=0.5cm, minimum height=0.55cm] (i) at (\ptreelfx + 4, \ptreelfy) {\textbf{a}};
    \node[draw, rectangle, minimum width=0.5cm, minimum height=0.55cm] (j) at (\ptreelfx + 4.5, \ptreelfy) {\textbf{g}};
    \node[draw, rectangle, minimum width=0.5cm, minimum height=0.55cm] (k) at (\ptreelfx + 5, \ptreelfy) {\textbf{c}};
    \node[draw, rectangle, minimum width=0.5cm, minimum height=0.55cm] (l) at (\ptreelfx + 5.5, \ptreelfy) {\textbf{g}};
    \node[draw, rectangle, minimum width=0.5cm, minimum height=0.55cm] (m) at (\ptreelfx + 6, \ptreelfy) {\textbf{c}};
    \node[draw, rectangle, minimum width=0.5cm, minimum height=0.55cm] (n) at (\ptreelfx + 6.5, \ptreelfy) {\textbf{g}};
    \node[draw, rectangle, minimum width=0.5cm, minimum height=0.55cm] (o) at (\ptreelfx + 7, \ptreelfy) {\textbf{c}};

    \node[draw, rectangle, minimum width=0.9cm, minimum height=0.45cm] (p) at (\ptreelflx + 0.5, \ptreelfly) {\textbf{$X_3$}};
    \node[inner sep=0pt, outer sep=0pt, minimum size=0pt] (1) at (\ptreelflx + 0.5, \ptreelfly - 0.75) {};
    \node[draw, rectangle, minimum width=0.9cm, minimum height=0.45cm] (q) at (\ptreelflx + 2, \ptreelfly) {\textbf{$X_2$}};
    \node[inner sep=0pt, outer sep=0pt, minimum size=0pt] (3) at (\ptreelflx + 2, \ptreelfly - 1.5) {};
    \node[draw, rectangle, minimum width=0.9cm, minimum height=0.45cm] (r) at (\ptreelflx + 3.5, \ptreelfly) {\textbf{$X_3$}};
    \node[inner sep=0pt, outer sep=0pt, minimum size=0pt] (2) at (\ptreelflx + 3.5, \ptreelfly - 0.75) {};
    \node[draw, rectangle, minimum width=0.9cm, minimum height=0.45cm] (s) at (\ptreelflx + 5, \ptreelfly) {\textbf{$X_2$}};
    \node[inner sep=0pt, outer sep=0pt, minimum size=0pt] (4) at (\ptreelflx + 5, \ptreelfly - 1.5) {};
    \node[draw, rectangle, minimum width=0.9cm, minimum height=0.45cm] (t) at (\ptreelflx + 6, \ptreelfly) {\textbf{$X_2$}};
    \node[inner sep=0pt, outer sep=0pt, minimum size=0pt] (5) at (\ptreelflx + 6, \ptreelfly - 1.5) {};
    
    \node[draw, rectangle, minimum width=0.9cm, minimum height=0.45cm] (v) at (\ptreelslx, \ptreelsly) {\textbf{$X_1$}};
    \node[inner sep=0pt, outer sep=0pt, minimum size=0pt] (6) at (\ptreelslx, \ptreelsly - 0.75) {};
    \node[draw, rectangle, minimum width=0.9cm, minimum height=0.45cm] (w) at (\ptreelslx + 1, \ptreelsly) {\textbf{$X_1$}};
    \node[inner sep=0pt, outer sep=0pt, minimum size=0pt] (7) at (\ptreelslx + 1, \ptreelsly - 0.75) {};
    \node[draw, rectangle, minimum width=0.9cm, minimum height=0.45cm] (x) at (\ptreelslx + 3, \ptreelsly) {\textbf{$X_1$}};
    \node[inner sep=0pt, outer sep=0pt, minimum size=0pt] (8) at (\ptreelslx + 3, \ptreelsly - 0.75) {};
    \node[draw, rectangle, minimum width=0.9cm, minimum height=0.45cm] (y) at (\ptreelslx + 4, \ptreelsly) {\textbf{$X_1$}};
    \node[inner sep=0pt, outer sep=0pt, minimum size=0pt] (9) at (\ptreelslx + 4, \ptreelsly - 0.75) {};

    \draw[-, line width=1.5pt, >=latex] (a) -- (10);
    \draw[->, line width=1.5pt, >=latex] (10) -| (p);
    \draw[->, line width=1.5pt, >=latex] (10) -| (q);
    \draw[->, line width=1.5pt, >=latex] (10) -- (r);
    \draw[->, line width=1.5pt, >=latex] (10) -| (s);
    \draw[->, line width=1.5pt, >=latex] (10) -| (t);
    \draw[->, line width=1.5pt, >=latex] (10) -| (o);
    \draw[-, line width=1.5pt, >=latex] (p) -- (1);
    \draw[->, line width=1.5pt, >=latex] (1) -| (v);
    \draw[->, line width=1.5pt, >=latex] (1) -| (w);
    \draw[-, line width=1.5pt, >=latex] (r) -- (2);
    \draw[->, line width=1.5pt, >=latex] (2) -| (x);
    \draw[->, line width=1.5pt, >=latex] (2) -| (y);
    \draw[-, line width=1.5pt, >=latex] (q) -- (3);
    \draw[->, line width=1.5pt, >=latex] (3) -| (e);
    \draw[->, line width=1.5pt, >=latex] (3) -| (f);
    \draw[-, line width=1.5pt, >=latex] (s) -- (4);
    \draw[->, line width=1.5pt, >=latex] (4) -| (k);
    \draw[->, line width=1.5pt, >=latex] (4) -| (l);
    \draw[-, line width=1.5pt, >=latex] (t) -- (5);
    \draw[->, line width=1.5pt, >=latex] (5) -| (m);
    \draw[->, line width=1.5pt, >=latex] (5) -| (n);

    \draw[-, line width=1.5pt, >=latex] (v) -- (6);
    \draw[->, line width=1.5pt, >=latex] (6) -| (b);
    \draw[->, line width=1.5pt, >=latex] (6) -| (c);
    \draw[-, line width=1.5pt, >=latex] (w) -- (7);
    \draw[->, line width=1.5pt, >=latex] (7) -| (d);
    \draw[->, line width=1.5pt, >=latex] (7) -| (dd);

    \draw[-, line width=1.5pt, >=latex] (x) -- (8);
    \draw[->, line width=1.5pt, >=latex] (8) -| (g);
    \draw[->, line width=1.5pt, >=latex] (8) -| (h);
    \draw[-, line width=1.5pt, >=latex] (y) -- (9);
    \draw[->, line width=1.5pt, >=latex] (9) -| (i);
    \draw[->, line width=1.5pt, >=latex] (9) -| (j);
    
\end{tikzpicture}

%% file: Figures/Panel_B.tex

\begin{tikzpicture}

    \newcommand{\ptreelzx}{-2.5}
    \newcommand{\ptreelzy}{0}
    \newcommand{\ptreelfx}{-6.25}
    \newcommand{\ptreelfy}{-4.5}
    \newcommand{\ptreelflx}{-6}
    \newcommand{\ptreelfly}{-1.5}
    \newcommand{\ptreelslx}{-6}
    \newcommand{\ptreelsly}{-3}
    
    \node[draw, rectangle, minimum width=0.9cm, minimum height=0.45cm] (a) at (\ptreelzx, \ptreelzy) {\textbf{$S$}};
    \node[inner sep=0pt, outer sep=0pt, minimum size=0pt] (10) at (\ptreelzx, \ptreelzy - 0.75) {};

    \node[draw, rectangle, minimum width=0.5cm, minimum height=0.55cm] (b) at (\ptreelfx + 0, \ptreelfy) {\textbf{a}};
    \node[draw, rectangle, minimum width=0.5cm, minimum height=0.55cm] (c) at (\ptreelfx + 0.5, \ptreelfy) {\textbf{g}};
    \node[draw, rectangle, minimum width=0.5cm, minimum height=0.55cm] (d) at (\ptreelfx + 1, \ptreelfy) {\textbf{a}};
    \node[draw, rectangle, minimum width=0.5cm, minimum height=0.55cm] (dd) at (\ptreelfx + 1.5, \ptreelfy) {\textbf{g}};
    \node[draw, rectangle, minimum width=0.5cm, minimum height=0.55cm] (e) at (\ptreelfx + 2, \ptreelfy) {\textbf{c}};
    \node[draw, rectangle, minimum width=0.5cm, minimum height=0.55cm] (f) at (\ptreelfx + 2.5, \ptreelfy) {\textbf{g}};
    \node[draw, rectangle, minimum width=0.5cm, minimum height=0.55cm] (g) at (\ptreelfx + 3, \ptreelfy) {\textbf{a}};
    \node[draw, rectangle, minimum width=0.5cm, minimum height=0.55cm] (h) at (\ptreelfx + 3.5, \ptreelfy) {\textbf{g}};
    \node[draw, rectangle, minimum width=0.5cm, minimum height=0.55cm] (i) at (\ptreelfx + 4, \ptreelfy) {\textbf{a}};
    \node[draw, rectangle, minimum width=0.5cm, minimum height=0.55cm] (j) at (\ptreelfx + 4.5, \ptreelfy) {\textbf{g}};
    \node[draw, rectangle, minimum width=0.5cm, minimum height=0.55cm] (k) at (\ptreelfx + 5, \ptreelfy) {\textbf{c}};
    \node[draw, rectangle, minimum width=0.5cm, minimum height=0.55cm] (l) at (\ptreelfx + 5.5, \ptreelfy) {\textbf{g}};
    \node[draw, rectangle, minimum width=0.5cm, minimum height=0.55cm] (m) at (\ptreelfx + 6, \ptreelfy) {\textbf{c}};
    \node[draw, rectangle, minimum width=0.5cm, minimum height=0.55cm] (n) at (\ptreelfx + 6.5, \ptreelfy) {\textbf{g}};
    \node[draw, rectangle, minimum width=0.5cm, minimum height=0.55cm] (o) at (\ptreelfx + 7, \ptreelfy) {\textbf{c}};

    \node[draw, rectangle, minimum width=0.9cm, minimum height=0.45cm] (p) at (\ptreelflx + 1, \ptreelfly) {\textbf{$X_3$}};
    \node[inner sep=0pt, outer sep=0pt, minimum size=0pt] (1) at (\ptreelflx + 1, \ptreelfly - 0.75) {};
    \node[draw, rectangle, minimum width=0.9cm, minimum height=0.45cm] (q) at (\ptreelflx + 4, \ptreelfly) {\textbf{$X_3$}};
    \node[inner sep=0pt, outer sep=0pt, minimum size=0pt] (2) at (\ptreelflx + 4, \ptreelfly - 0.75) {};
    \node[draw, rectangle, minimum width=0.9cm, minimum height=0.45cm] (r) at (\ptreelflx + 6, \ptreelfly) {\textbf{$X_2$}};
    \node[inner sep=0pt, outer sep=0pt, minimum size=0pt] (3) at (\ptreelflx + 6, \ptreelfly - 1.5) {};

     \node[draw, rectangle, minimum width=0.9cm, minimum height=0.45cm] (s) at (\ptreelslx + 2, \ptreelsly) {\textbf{$X_2$}};
    \node[inner sep=0pt, outer sep=0pt, minimum size=0pt] (4) at (\ptreelslx + 2, \ptreelsly - 0.75) {};
    \node[draw, rectangle, minimum width=0.9cm, minimum height=0.45cm] (t) at (\ptreelslx + 5, \ptreelsly) {\textbf{$X_2$}};
    \node[inner sep=0pt, outer sep=0pt, minimum size=0pt] (5) at (\ptreelslx + 5, \ptreelsly - 0.75) {};
    \node[draw, rectangle, minimum width=0.9cm, minimum height=0.45cm] (v) at (\ptreelslx, \ptreelsly) {\textbf{$X_1$}};
    \node[inner sep=0pt, outer sep=0pt, minimum size=0pt] (6) at (\ptreelslx, \ptreelsly - 0.75) {};
    \node[draw, rectangle, minimum width=0.9cm, minimum height=0.45cm] (w) at (\ptreelslx + 1, \ptreelsly) {\textbf{$X_1$}};
    \node[inner sep=0pt, outer sep=0pt, minimum size=0pt] (7) at (\ptreelslx + 1, \ptreelsly - 0.75) {};
    \node[draw, rectangle, minimum width=0.9cm, minimum height=0.45cm] (x) at (\ptreelslx + 3, \ptreelsly) {\textbf{$X_1$}};
    \node[inner sep=0pt, outer sep=0pt, minimum size=0pt] (8) at (\ptreelslx + 3, \ptreelsly - 0.75) {};
    \node[draw, rectangle, minimum width=0.9cm, minimum height=0.45cm] (y) at (\ptreelslx + 4, \ptreelsly) {\textbf{$X_1$}};
    \node[inner sep=0pt, outer sep=0pt, minimum size=0pt] (9) at (\ptreelslx + 4, \ptreelsly - 0.75) {};

    \draw[-, line width=1.5pt, >=latex] (a) -- (10);
    \draw[->, line width=1.5pt, >=latex] (10) -| (p);
    \draw[->, line width=1.5pt, >=latex] (10) -| (q);
    \draw[->, line width=1.5pt, >=latex] (10) -| (r);
    \draw[->, line width=1.5pt, >=latex] (10) -| (o);
    \draw[-, line width=1.5pt, >=latex] (p) -- (1);
    \draw[->, line width=1.5pt, >=latex] (1) -| (v);
    \draw[->, line width=1.5pt, >=latex] (1) -- (w);
    \draw[->, line width=1.5pt, >=latex] (1) -| (s);
    \draw[-, line width=1.5pt, >=latex] (q) -- (2);
    \draw[->, line width=1.5pt, >=latex] (2) -| (x);
    \draw[->, line width=1.5pt, >=latex] (2) -- (y);
    \draw[->, line width=1.5pt, >=latex] (2) -| (t);
    \draw[-, line width=1.5pt, >=latex] (r) -- (3);
    \draw[->, line width=1.5pt, >=latex] (3) -| (n);
    \draw[->, line width=1.5pt, >=latex] (3) -| (m);

    \draw[-, line width=1.5pt, >=latex] (s) -- (4);
    \draw[->, line width=1.5pt, >=latex] (4) -| (f);
    \draw[->, line width=1.5pt, >=latex] (4) -| (e);
    \draw[-, line width=1.5pt, >=latex] (t) -- (5);
    \draw[->, line width=1.5pt, >=latex] (5) -| (k);
    \draw[->, line width=1.5pt, >=latex] (5) -| (l);

    \draw[-, line width=1.5pt, >=latex] (v) -- (6);
    \draw[->, line width=1.5pt, >=latex] (6) -| (b);
    \draw[->, line width=1.5pt, >=latex] (6) -| (c);
    \draw[-, line width=1.5pt, >=latex] (w) -- (7);
    \draw[->, line width=1.5pt, >=latex] (7) -| (d);
    \draw[->, line width=1.5pt, >=latex] (7) -| (dd);
    \draw[-, line width=1.5pt, >=latex] (x) -- (8);
    \draw[->, line width=1.5pt, >=latex] (8) -| (g);
    \draw[->, line width=1.5pt, >=latex] (8) -| (h);

    \draw[-, line width=1.5pt, >=latex] (x) -- (8);
    \draw[->, line width=1.5pt, >=latex] (8) -| (g);
    \draw[->, line width=1.5pt, >=latex] (8) -| (h);
    \draw[-, line width=1.5pt, >=latex] (y) -- (9);
    \draw[->, line width=1.5pt, >=latex] (9) -| (i);
    \draw[->, line width=1.5pt, >=latex] (9) -| (j);
    
\end{tikzpicture}

%% file: Figures/Panel_C.tex

\begin{tikzpicture}

    \newcommand{\bvectx}{-5}
    \newcommand{\bvecty}{0}
    \newcommand{\rulex}{-2.25}
    \newcommand{\ruley}{0}
    \newcommand{\sizex}{-4.25}
    \newcommand{\sizey}{0}
    \newcommand{\charx}{-3.5}
    \newcommand{\chary}{0}
    \newcommand{\textx}{0.5}
    \newcommand{\texty}{-0.6}
    
    \node[rectangle, minimum width=0.65cm, minimum height=0.6cm] (a) at (\bvectx, \bvecty) {\textbf{$B_X$}};
    \node[draw, rectangle, minimum width=0.65cm, minimum height=0.6cm] (b) at (\bvectx, \bvecty - 0.6) {\textbf{1}};
    \node[draw, rectangle, minimum width=0.65cm, minimum height=0.6cm] (c) at (\bvectx, \bvecty - 1.2) {\textbf{0}};
    \node[draw, rectangle, minimum width=0.65cm, minimum height=0.6cm] (d) at (\bvectx, \bvecty - 1.8) {\textbf{1}};
    \node[draw, rectangle, minimum width=0.65cm, minimum height=0.6cm] (e) at (\bvectx, \bvecty - 2.4) {\textbf{1}};

    \node[rectangle, minimum width=1.5cm, minimum height=0.6cm] (a) at (\rulex, \ruley) {$X$};
    \node[draw, rectangle, minimum width=1.0cm, minimum height=0.6cm] (b) at (\rulex, \ruley - 0.6) {\makebox[1.5cm][l]{\textbf{ag}}};
    \node[draw, rectangle, minimum width=1.0cm, minimum height=0.6cm] (c) at (\rulex, \ruley - 1.2) {\makebox[1.5cm][l]{\textbf{cg}}};
    \node[draw, rectangle, minimum width=1.0cm, minimum height=0.6cm] (d) at (\rulex, \ruley - 1.8) {\makebox[1.5cm][l]{\textbf{$X_1X_1X_2$}}};
    \node[draw, rectangle, minimum width=1.0cm, minimum height=0.6cm] (e) at (\rulex, \ruley - 2.4) {\makebox[1.5cm][l]{\textbf{$X_3X_3X_2$c}}};

    \node[rectangle, minimum width=0.65cm, minimum height=0.6cm] (a) at (\sizex, \sizey) {\textbf{$L$}};
    \node[draw, rectangle, minimum width=0.65cm, minimum height=1.2cm] (b) at (\sizex, \sizey - 0.9) {\textbf{2}};
    \node[draw, rectangle, minimum width=0.65cm, minimum height=0.6cm] (d) at (\sizex, \sizey - 1.8) {\textbf{6}};
    \node[draw, rectangle, minimum width=0.65cm, minimum height=0.6cm] (e) at (\sizex, \sizey - 2.4) {\textbf{15}};

    \node[rectangle, minimum width=0.6cm, minimum height=0.6cm] (b) at (\charx, \chary - 0.6) {\textbf{$X_1$}};
    \node[rectangle, minimum width=0.6cm, minimum height=0.6cm] (c) at (\charx, \chary - 1.2) {\textbf{$X_2$}};
    \node[rectangle, minimum width=0.6cm, minimum height=0.6cm] (d) at (\charx, \chary - 1.8) {\textbf{$X_3$}};
    \node[rectangle, minimum width=0.6cm, minimum height=0.6cm] (e) at (\charx, \chary - 2.4) {\textbf{$S$}};

     \node[rectangle, minimum width=0.5cm, minimum height=0.6cm] (zz) at (\textx - 0.6, \texty) {\textbf{$T=$}};
    \node[rectangle, minimum width=0.5cm, minimum height=0.55cm] (b) at (\textx + 0, \texty) {\textbf{a}};
    \node[rectangle, minimum width=0.5cm, minimum height=0.55cm] (c) at (\textx + 0.5, \texty) {\textbf{g}};
    \node[rectangle, minimum width=0.5cm, minimum height=0.55cm] (d) at (\textx + 1, \texty) {\textbf{a}};
    \node[rectangle, minimum width=0.5cm, minimum height=0.55cm] (dd) at (\textx + 1.5, \texty) {\textbf{g}};
    \node[rectangle, minimum width=0.5cm, minimum height=0.55cm] (e) at (\textx + 2, \texty) {\textbf{c}};
    \node[rectangle, minimum width=0.5cm, minimum height=0.55cm] (f) at (\textx + 2.5, \texty) {\textbf{g}};
    \node[rectangle, minimum width=0.5cm, minimum height=0.55cm] (g) at (\textx + 3, \texty) {\textbf{a}};
    \node[rectangle, minimum width=0.5cm, minimum height=0.55cm] (h) at (\textx + 3.5, \texty) {\textbf{g}};
    \node[rectangle, minimum width=0.5cm, minimum height=0.55cm] (i) at (\textx + 4, \texty) {\textbf{a}};
    \node[rectangle, minimum width=0.5cm, minimum height=0.55cm] (j) at (\textx + 4.5, \texty) {\textbf{g}};
    \node[rectangle, minimum width=0.5cm, minimum height=0.55cm] (k) at (\textx + 5, \texty) {\textbf{c}};
    \node[rectangle, minimum width=0.5cm, minimum height=0.55cm] (l) at (\textx + 5.5, \texty) {\textbf{g}};
    \node[rectangle, minimum width=0.5cm, minimum height=0.55cm] (m) at (\textx + 6, \texty) {\textbf{c}};
    \node[rectangle, minimum width=0.5cm, minimum height=0.55cm] (n) at (\textx + 6.5, \texty) {\textbf{g}};
    \node[rectangle, minimum width=0.5cm, minimum height=0.55cm] (o) at (\textx + 7, \texty) {\textbf{c}};

    \node[rectangle, minimum width=0.5cm, minimum height=0.6cm] (ba) at (\textx - 0.8, \texty - 1.05) {\textbf{$S$}};
    \node[minimum width=0.5cm, minimum height=0.6cm] (bz) at (\textx, \texty - 1.05) {};
    \draw[->, line width=0.75pt, >=latex] (ba) -- (bz);

    \newcommand{\mvoff}{0.3}
    \node[rectangle, minimum width=0.5cm, minimum height=0.55cm, font=\small] (bb) at (\textx + 0, \texty - \mvoff) {1};
    \node[rectangle, minimum width=0.5cm, minimum height=0.55cm, font=\small] (bc) at (\textx + 0.5, \texty - \mvoff) {2};
    \node[rectangle, minimum width=0.5cm, minimum height=0.55cm, font=\small] (bd) at (\textx + 1, \texty - \mvoff) {3};
    \node[rectangle, minimum width=0.5cm, minimum height=0.55cm, font=\small] (be) at (\textx + 1.5, \texty - \mvoff) {4};
    \node[rectangle, minimum width=0.5cm, minimum height=0.55cm, font=\small] (bf) at (\textx + 2, \texty - \mvoff) {5};
    \node[rectangle, minimum width=0.5cm, minimum height=0.55cm, font=\small] (bg) at (\textx + 2.5, \texty - \mvoff) {6};
    \node[rectangle, minimum width=0.5cm, minimum height=0.55cm, font=\small] (bh) at (\textx + 3, \texty - \mvoff) {7};
    \node[rectangle, minimum width=0.5cm, minimum height=0.55cm, font=\small] (bi) at (\textx + 3.5, \texty - \mvoff) {8};
    \node[rectangle, minimum width=0.5cm, minimum height=0.55cm, font=\small] (bj) at (\textx + 4, \texty - \mvoff) {9};
    \node[rectangle, minimum width=0.5cm, minimum height=0.55cm, font=\small] (bk) at (\textx + 4.5, \texty - \mvoff) {10};
    \node[rectangle, minimum width=0.5cm, minimum height=0.55cm, font=\small] (bl) at (\textx + 5, \texty - \mvoff) {11};
    \node[rectangle, minimum width=0.5cm, minimum height=0.55cm, font=\small] (bm) at (\textx + 5.5, \texty - \mvoff) {12};
    \node[rectangle, minimum width=0.5cm, minimum height=0.55cm, font=\small] (bn) at (\textx + 6, \texty - \mvoff) {13};
    \node[rectangle, minimum width=0.5cm, minimum height=0.55cm, font=\small] (bo) at (\textx + 6.5, \texty - \mvoff) {14};
    \node[rectangle, minimum width=0.5cm, minimum height=0.55cm, font=\small] (bp) at (\textx + 7, \texty - \mvoff) {15};

    \newcommand{\voff}{-1.75}
    \node[rectangle, minimum width=0.5cm, minimum height=0.6cm] (az) at (\textx - 0.75, \texty + \voff) {\textbf{$B_S$}};
    \node[draw, rectangle, minimum width=0.5cm, minimum height=0.55cm] (p) at (\textx + 0, \texty + \voff) {\textbf{1}};
    \node[draw, rectangle, minimum width=0.5cm, minimum height=0.55cm] (q) at (\textx + 0.5, \texty + \voff) {\textbf{0}};
    \node[draw, rectangle, minimum width=0.5cm, minimum height=0.55cm] (r) at (\textx + 1, \texty + \voff) {\textbf{0}};
    \node[draw, rectangle, minimum width=0.5cm, minimum height=0.55cm] (s) at (\textx + 1.5, \texty + \voff) {\textbf{0}};
    \node[draw, rectangle, minimum width=0.5cm, minimum height=0.55cm] (t) at (\textx + 2, \texty + \voff) {\textbf{0}};
    \node[draw, rectangle, minimum width=0.5cm, minimum height=0.55cm] (u) at (\textx + 2.5, \texty + \voff) {\textbf{0}};
    \node[draw, rectangle, minimum width=0.5cm, minimum height=0.55cm] (v) at (\textx + 3, \texty + \voff) {\textbf{1}};
    \node[draw, rectangle, minimum width=0.5cm, minimum height=0.55cm] (w) at (\textx + 3.5, \texty + \voff) {\textbf{0}};
    \node[draw, rectangle, minimum width=0.5cm, minimum height=0.55cm] (x) at (\textx + 4, \texty + \voff) {\textbf{0}};
    \node[draw, rectangle, minimum width=0.5cm, minimum height=0.55cm] (y) at (\textx + 4.5, \texty + \voff) {\textbf{0}};
    \node[draw, rectangle, minimum width=0.5cm, minimum height=0.55cm] (z) at (\textx + 5, \texty + \voff) {\textbf{0}};
    \node[draw, rectangle, minimum width=0.5cm, minimum height=0.55cm] (aa) at (\textx + 5.5, \texty + \voff) {\textbf{0}};
    \node[draw, rectangle, minimum width=0.5cm, minimum height=0.55cm] (ab) at (\textx + 6, \texty + \voff) {\textbf{1}};
    \node[draw, rectangle, minimum width=0.5cm, minimum height=0.55cm] (ac) at (\textx + 6.5, \texty + \voff) {\textbf{0}};
    \node[draw, rectangle, minimum width=0.5cm, minimum height=0.55cm] (ad) at (\textx + 7, \texty + \voff) {\textbf{1}};

    \newcommand{\vboff}{-0.3}
    \draw[line width=1pt] (\textx - 0.15,\texty + \vboff - 0.3) -- (\textx + 2.5 + 0.15,\texty + \vboff - 0.3) node[midway,below=6pt] {\textbf{$X_3$}};
    \draw[line width=1pt] (\textx + 1.25,\texty + \vboff - 0.3) -- (\textx + 1.25,\texty + \vboff - 0.45);
    \draw[line width=1pt] (\textx - 0.15,\texty + \vboff - 0.3) -- (\textx -0.15,\texty + \vboff - 0.15);
    \draw[line width=1pt] (\textx + 2.5 + 0.15,\texty + \vboff - 0.3) -- (\textx + 2.5 + 0.15,\texty + \vboff - 0.15);
    \draw[line width=1pt] (\textx  + 3 - 0.15,\texty + \vboff - 0.3) -- (\textx + 5.5 + 0.15,\texty + \vboff - 0.3) node[midway,below=6pt] {\textbf{$X_3$}};
    \draw[line width=1pt] (\textx + 3 + 1.25,\texty + \vboff - 0.3) -- (\textx + 3 + 1.25,\texty + \vboff - 0.45);
    \draw[line width=1pt] (\textx + 3 - 0.15,\texty + \vboff - 0.3) -- (\textx + 3 - 0.15,\texty + \vboff - 0.15);
    \draw[line width=1pt] (\textx + 5.5 + 0.15,\texty + \vboff - 0.3) -- (\textx + 5.5 + 0.15,\texty + \vboff - 0.15);
    \draw[line width=1pt] (\textx  + 6 - 0.15,\texty + \vboff - 0.3) -- (\textx + 6.5 + 0.15,\texty + \vboff - 0.3) node[midway,below=6pt] {\textbf{$X_2$}};
    \draw[line width=1pt] (\textx + 5 + 1.25,\texty + \vboff - 0.3) -- (\textx + 5 + 1.25,\texty + \vboff - 0.45);
    \draw[line width=1pt] (\textx + 6 - 0.15,\texty + \vboff - 0.3) -- (\textx + 6 - 0.15,\texty + \vboff - 0.15);
    \draw[line width=1pt] (\textx + 6.5 + 0.15,\texty + \vboff - 0.3) -- (\textx + 6.5 + 0.15,\texty + \vboff - 0.15);
    \draw[line width=1pt] (\textx  + 7 - 0.15,\texty + \vboff - 0.3) -- (\textx + 7 + 0.15,\texty + \vboff - 0.3) node[midway,below=8pt] {\textbf{c}};
    \draw[line width=1pt] (\textx + 7,\texty + \vboff - 0.3) -- (\textx + 7,\texty + \vboff - 0.45);
    \draw[line width=1pt] (\textx + 7 - 0.15,\texty + \vboff - 0.3) -- (\textx + 7 - 0.15,\texty + \vboff - 0.15);
    \draw[line width=1pt] (\textx + 7 + 0.15,\texty + \vboff - 0.3) -- (\textx + 7 + 0.15,\texty + \vboff - 0.15);
    
\end{tikzpicture}